\documentclass[3p,times,sort&compress]{elsarticle}
\usepackage{hyperref}

\usepackage{lineno}

\usepackage{amssymb,amsmath,graphicx}
\usepackage{epigraph}

\begin{document}

\begin{frontmatter}

\title{Singularities of transient processes in dynamics and beyond}

\author{A.N. Gorban}
\ead{a.n.gorban@le.ac.uk}

\address{Department of Mathematics, University of Leicester, Leicester, LE1 7RH, UK}

\begin{abstract}
This note is a brief review of the analysis of long transients in dynamical systems. The problem of long transients arose in many disciplines, from physical and chemical kinetic to biology and even social sciences. Detailed analysis of singularities of various `relaxation times' associated long transients with bifurcations of $\omega$-limit sets, homoclinic structures (intersections of $\alpha$- and $\omega$-limit sets) and other peculiarities of dynamics. This review was stimulated by the analysis of anomalously long transients in ecology published recently by A. Morozov and S. Petrovskii with co-authors.
\end{abstract}
     
\end{frontmatter}
\vspace{5mm}

\epigraph{Sic Transit Gloria Mundi}{STGM - a famous Latin phrase}
	
From an absolutely rigorous (pedantic) point of view, 	all the states in life or social sciences (and, possibly, far beyond their borders) are transient. All the  steady states, running waves, beautiful limit cycles or attractors are just {\em intermediate asymptotics} and nothing is stationary or ergodic. 

However, the idea of {\em separation of  time scales} allows the  creation of autonomous dynamic models at some scales. In these ideal models, we assume that processes that are much faster than the selected time scale are completely relaxed and the values of fast variables (or average values in fast dynamics) follow the dynamics at the selected scale. The processes proceeding much more slowly than the selected time can be considered stationary or presented as a slow drift of the parameters of our models. This clear and transparent picture is true if the system is globally stable and far from a critical transition. Such a system relaxes to the limit regime without significant delays, and all long transients are due to a slow drift of parameters.
When the qualitative picture of dynamics is not so trivial, then various dynamic causes of slow relaxations and critical delay effects, well-known in physics and beyond, may appear (we refer here to a recent review of {\em critical transitions} \cite{AnticipatingScience}). Critical transitions can {\em mix time scales} and violate the standard separation of time logics. Even violation of global stability can cause long delay near unstable attractors.

If we observe a long transition period for a real system, a conundrum arises: 
\begin{itemize}
\item {\em is this delay caused by the drift of ``external'' conditions (parameters),}
\item {\em or does it have an internal dynamic cause,}
\item {\em or does it appear just due to the inaccuracy of the model, because we incorrectly determine the limit behaviour and transients?}
\end{itemize}
 This problem can be very non-trivial when the barrier between reality and models is large enough, as in ecology (and, more broadly, in   life science), in heterogeneous catalysis, the dynamics of complex solids and liquids, etc., where we do not expect very high precision models.

Recently published papers  \cite{MorozovScience,MorozovPLRV} met this challenge for mathematical modelling in ecology. I liked this study and review with a clear presentation of the main ideas and results important for ecological research. In this comment, I would like to enrich the discussion with a similar but different story from chemical kinetics and the introduction of general mathematical concepts developed some time ago for the analysis of long transients in kinetics and dynamical systems.

Transients have been used  in experimental chemistry to measure reaction rate constants, especially for fast reactions (M. Eigen, Nobel prize 1967 \cite{EIgenNobel}). The study of transients attracted much attention, and slow transients were observed in the kinetics of heterogeneous catalytic reactions, which were immediately interpreted as a drift of `external' conditions \cite{Temkin1,Temkin2,Wainwright}. In the same period of time, chemical kinetics mastered the achievements of non-linear dynamics with dramatic discoveries and rediscoveries of chemical oscillations, stochastic self-oscillations, auto-waves, etc. {\em When the problem of long transients  met with the theory of dynamical systems, it immediately became clear that it is not always necessary to look for `external' causes of slow relaxations, first of all it is necessary to investigate whether there are slow relaxations of dynamic origin in the system} \cite{GorbanDiss,GorbanDAN} (a slightly abridged version of the thesis \cite{GorbanDiss} is published in English \cite{GorbanSingularities}; detailed discussion of long transients in chemical kinetics and the theory of their possible dynamic causes is presented in Chapter 7, `Critical retardation effects and slow relaxations', of \cite{YBGE1991}).

Before mathematical analysis of long transients, two questions should be answered: what we call the {\em limiting behaviour} for the time $ t \to \infty $ and how we define the {\em relaxation time}. For asymptotically stable linear systems $\dot{x}=Ax$,  relaxation time is traditionally defined as $ \tau_l = -1/\max {\rm Re} \lambda$ where $\lambda$ runs through all  the  eigenvalues of the matrix  $A$ (all $ {\rm Re} \lambda <0$). For non-linear systems, this definition makes sense only for transients in small neighbourhoods of asymptotically stable equilibria, but not for global dynamics far from such equilibria.

Non-linear dynamics gives a nice formalisation of limit behaviour, the {\em $\omega$-limit set}. For a motion $x(t)$ the $\omega$-limit set consists of all limit points of $x(t)$ when $t\to \infty$. For theory of $\alpha$- (when $t \to \infty$) and $\omega$-limit sets we refer to \cite{GorbanSingularities} or to Chapter VII, `General theory of dynamical system', of the classical monograph  \cite{Birkhoff1927}. If a system depends on parameters $k$ then the $\omega$-limit set can be considered as a set-valued function of parameters $k$ and initial state of the motion $x$: $\omega(x,k)$. If a motion goes to an equilibrium then its $\omega$-limit set consists of this equilibrium. If it goes to a periodic orbit then the  $\omega$-limit set consists of the points of this orbit. Of course, $\omega(x,k)$  includes point of more complex attractors if the motion tends to them.

{\em Transient is the process of relaxation from the initial state to the small $\varepsilon$-vicinity of $\omega(x,k)$}. The relaxation time depends on $x$, $k$, and $\varepsilon>0$. The value of $\varepsilon$, `accuracy of relaxation', is to be fixed because for smaller $\varepsilon$ relaxation will go longer, and for unlimited accuracy, when $\varepsilon \to 0$, the relaxation time goes to infinity. Therefore, we consider relaxation time as a function of $x$ and $k$ for a given $\varepsilon$. If we go deeper then we notice that, after the first entrance into the  $\varepsilon$-vicinity of $\omega(x,k)$, the motion can leave this vicinity, then return back, then, possibly leave it again, etc., and the time of the final entrance into  $\varepsilon$-vicinity of $\omega(x,k)$ could be much larger than the time of the first entrance. Therefore, we can define {\em several relaxation times}: 
\begin{enumerate}
\item time of the first entrance into the  $\varepsilon$-vicinity of $\omega(x,k)$; 
\item time of the motion outside it; 
\item time of the final entrance there.
\end{enumerate} 
There are several other types of relaxation time \cite{GorbanSingularities}. Therefore, a large number of different slow relaxations arise, not reducible to each other. To interpret the long-term transient observed experimentally, it is important to understand which of the relaxation times is long.

Long transients in experiments and computational experiments are usually {\em `limited slow'}, the relaxation time is large (larger than one would expect from the coefficients of equations and  of characteristic times of elementary processes), but nevertheless limited (for given $\varepsilon$). How can `long transients' be separated from the `normal' ones?  To find singularities of relaxation time, the following method is useful, which goes back to the works of A.A. Andronov: the system in question is included in an appropriate family of dynamical systems for which relaxation times already have singularities (are not bounded). These singularities of transients appear when the function $\omega(x,k)$ looses its continuity. More precisely, singularities of relaxation time are caused by the  `$\omega(x,k)$ explosions', that are the discontinuities, where new $\omega$-limit points or whole $\omega$-limit sets appear \cite{GorbanDiss,GorbanDAN,GorbanSingularities}. The second common reason of anomalously long transients is appearance of homoclinic structures, that are intersections between $\omega$- ($t\to \infty$) and $\alpha$- ($t\to -\infty$) limit sets.

The surface of singularities of relaxation time in the $(x,k)$ space includes the sets of attraction of unstable invariant sets. Examples of such piece-wise differentiable sets were calculated for some chemical systems \cite{BykGorPush1982} (see also \cite{YBGE1991}). 

\begin{figure}
\centering
\includegraphics[width=0.3\textwidth]{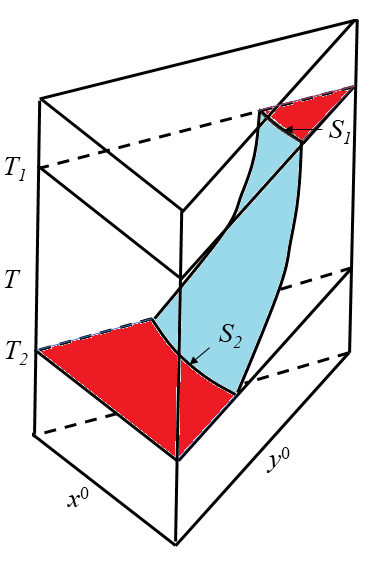}
\caption{{\em Singularities of relaxation time for catalytic trigger.} The surface of singularities of relaxation time is drawn in 3D space: dynamical variables $x$, $y$ and parameter $T$. Relaxation time depends on the initial conditions $x^0$, $y^0$ and on the value of parameter $T$. For the critical values $T=T_{1,2}$ the system has the saddle-node bifurcations. The basins of attraction of the saddle-nodes belong to the surface of singularities (highlighted in red). In between the critical values, the surface of singularities is formed by separatrices of saddles (highlighted in blue). Separatrices $S_{1,2}$ separate in phase space basins of attractors of saddle-nodes from the basins of attractors of stable nodes.}
\label{Fig:SingSurf}
\end{figure}

In Fig.~\ref{Fig:SingSurf}, we present the surface of singularities of relaxation time for the catalytic trigger, a simplest catalytic reaction without autocatalysis that allows multiplicity of steady states.
\begin{eqnarray*}
&{A_2}+2Z \leftrightarrow 2AZ; \\
&{B}+Z \leftrightarrow BZ;\\
&{AZ}+BZ \to {AB}+2Z.
\end{eqnarray*}
Here, $A_2$, $B$ and $AB$  are gases (for example, O$_2$, CO and CO$_2$), $Z$ is the ``adsorption place'' on the surface of the solid catalyst (for example, Pt), $AZ$ and $BZ$ are the intermediates on the surface. Dynamics of the intermediates on the surface was studied for  constant gas pressures \cite{YBGE1991,BykGorPush1982}. For different parameters (gas pressures and temperature), the system has either one steady state (a stable node), three steady states (two stable nodes and a saddle point) or two steady states at the bifurcations between one and three states: one stable node and  and one saddle-node. The surface of singularities of relaxation time is drawn (Fig.~\ref{Fig:SingSurf}) in the 3D space with coordinates: $x$ -- surface concentration of $AZ$, $y$ -- surface concentration of $BZ$ and $T$ -- temperature. The rate constants and their dependence on temperature correspond to the oxidation of CO on platinum at low pressure. The saddle-node bifurcation occurs at two values of the parameter $T=T_{1,2}$ \cite{YBGE1991,BykGorPush1982}. For these values, the singularity surface consists of the basins of attractors of the saddle-node point, and for $T$ between the bifurcation values, this surface is the union of the separatrices of the saddles. At a critical value $T=T_{1,2}$, the separatrix $S_{1,2}$ separates the basin  of attraction of the saddle-node  from the basin of attraction of the stable node. 

It could be curious to note that the theory of slow transients and bifurcations of $\omega$-limit sets \cite{GorbanDiss,GorbanDAN} (see also \cite{GorbanSingularities}) was created in topological dynamics (1978-1981) even before Milnor's introduction (1985) of the concept of `attractor' \cite{Milnor1985}.

Despite some differences at the level of formalism, at the {\em qualitative level}, the conclusions of the theory of dynamical systems on critical delays \cite{GorbanDiss,GorbanDAN,GorbanSingularities}, the results of the analysis of chemical kinetics \cite{YBGE1991} and the analysis of long-term transients in ecology \cite{MorozovScience,MorozovPLRV} {\em led to very similar conclusions}. For example:
\begin{itemize}
\item{It is not always necessary to search for `external' reasons of slow relaxations, in the first
place one should  investigate  if there are  slow  relaxations  of  dynamical origin in the system.}
\item{One of possible reasons for slow relaxations is the existence  of bifurcations (explosions) of $\omega$-limit sets $\omega(x,k)$ \cite{GorbanDiss,GorbanDAN,GorbanSingularities}. Particular cases are delays near unstable invariant sets and `ghost' attractors that will approach the real phase space with changing  parameters (for example, impact of non-physical steady states on transients was demonstrated for chemical kinetic systems in \cite{GorbanNonPhys}).}
\item{The measure (volume, probability) of the set of the initial conditions with large relaxation time can be asymptotically estimated using the spectrum of Lyapunov exponents of unstable invariant sets and ghost attractors. }
\item{The complicated dynamics can be `coarsed' by perturbations. The useful model of perturbations in topological dynamics provide the $\varepsilon$-motions ($\varepsilon$-orbits or pseudo-orbits \cite{GorbanSingularities,Corless1995,GorbanCoarseGrain}). For $\varepsilon \rightarrow 0$ we obtain the coarse structure of sources and drains similar to the Morse-Smale systems (possibly, with a totally disconnected compact instead of the finite set of attractors).}
\item{The interrelations between the singularities of relaxation
times and other peculiarities of dynamics for general dynamical system under small perturbations are the same as for the Morse-Smale systems, and, in particular, the same as for rough (structurally stable) two-dimensional systems \cite{GorbanDiss,GorbanSingularities}.}
\end{itemize}

This convergence of knowledge convinces us that the truth is not far away. Let us study singularities of transients! They make much sense and help us to anticipate critical effects.

 \end{document}